# On some experimental features of car-following behavior and how to model them

Rui Jiang[1,2], Mao-Bin Hu[2], H.M.Zhang[3,4], Zi-You Gao[1], Bin Jia[1], Qing-Song Wu[2]

[1]MOE Key Laboratory for Urban Transportation Complex Systems Theory and Technology, Beijing Jiaotong University, Beijing 100044, China

[2] State Key Laboratory of Fire Science and School of Engineering Science, University of Science and Technology of China, Hefei 230026, China

[3]Department of Civil and Environmental Engineering, University of California Davis, Davis, California 95616, United States of America

[4]Department of Traffic Engineering, School of Transportation Engineering, Tongji University, Shanghai 200092, China

**Abstract:** We have carried out car-following experiments with a 25-car-platoon on an open road section to study the relation between a car's speed and its spacing under various traffic conditions, in the hope to resolve a controversy surrounding this fundamental relation of vehicular traffic. In this paper we extend our previous analysis of these experiments, and report new experimental findings. In particular, we reveal that the platoon length (hence the average spacing within a platoon) might be significantly different even if the average velocity of the platoon is essentially the same. The findings further demonstrate that the traffic states span a 2D region in the speed-spacing (or density) plane. The common practice of using a single speed-spacing curve to model vehicular traffic ignores the variability and imprecision of human driving and is therefore inadequate. We have proposed a car-following model based on a mechanism that in certain ranges of speed and spacing, drivers are insensitive to the changes in spacing when the velocity differences between cars are small. It was shown that the model can reproduce the experimental results well.

## 1. Introduction

Uninterrupted traffic flows on highways, freeways, or expressways exhibit many complex phenomena such as the formation of phantom jam, the capacity drop, the wide scattering of data in the flow-density plane in congested traffic, and many traffic flow models have been proposed to simulate traffic flow dynamics (see e.g., Chowdhury et al., 2000; Helbing, 2001; Nagel et al., 2003 and references therein). However, up to now, the physics of traffic flow has not been fully understood and there are still heated debates about certain aspects of traffic flow theory, for example, the validity of Kerner's two-dimensional flow-density hypothesis (Schönhof and Helbing, 2007, 2009; Treiber et al., 2010; Kerner, 2013).

Traffic flow complexity comes from both car-following and lane-changing behavior. In microscopic traffic flow research, one usually first establishes and studies the car-following models, then introduces lane changing behavior and other measurements into the car-following models to deal with multilane traffic flow, mixed traffic flow, on-ramps, off-ramps, etc. Therefore, understanding car-following behavior forms the basis of and plays a vital role in developing micro traffic models.

In the classical General-Motors (GM) family of car-following models (Chandler et al., 1958), the acceleration of a vehicle $n$ is proportional to the velocity difference to the preceding one $n-1$

$$\frac{dv_n(t+\tau)}{dt} = \lambda[v_{n-1}(t) - v_n(t)] \tag{1}$$

The stability of traffic flow is related to the sensitivity $\lambda$ and the reaction time $\tau$: traffic flow is unstable when $\lambda\tau > 1/2$, which is used to describe the formation of jams. Here $v$ is velocity of vehicles. This GM model has later been extended to take into account the effects of spacing and velocity of the following vehicle on acceleration (Gazis et al., 1959, 1961; Herman et al., 1959):

$$\frac{dv_n(t+\tau)}{dt} = \lambda(v_n(t+\tau))^m [v_{n-1}(t) - v_n(t)]/[x_{n-1}(t) - x_n(t)]^l \tag{2}$$

where $m$ and $l$ are two parameters. A homogeneous traffic state is unstable when $\frac{\lambda\tau v^m}{(\Delta x)^l} > 1/2$. Here $\Delta x$ is the spacing between vehicles.

Different from the GM models, Newell (1961) developed a model in which it is assumed that there is a unique relationship $V(\Delta x)$ between speed $v$ and spacing $\Delta x$, and the velocity is determined by

$$v_n(t+\tau) = V(\Delta x(t)) \tag{3}$$

Making Taylor's expansion and reformulating (3), one could derive

$$\frac{dv_n(t)}{dt} = [V(\Delta x(t)) - v_n(t)]/\tau \tag{4}$$

which is another well known model, the so-called optimal velocity (OV) model (Bando et al., 1995). Traffic in this model is unstable when

$$V'(\Delta x) > 1/(2\tau) \tag{5}$$

In the OV model, Bando et al. (1995) used an OV function with a turning point[1]

$$V(\Delta x) = \frac{v_{max}}{2}\left[\tanh(\Delta x - h) + \tanh(h)\right] \tag{6}$$

Thus, traffic flow is stable in the high and low density range but unstable in the intermediate density range. Therefore, it has been claimed that the OV model is able to describe the formation of stop-start waves.

In many other car-following models, such as the intelligent driver (ID) model (Treiber et al., 2000), the full velocity difference (FVD) model (Jiang et al., 2001), it is also assumed either explicitly or implicitly that there is a unique relationship $V(\Delta x)$ between speed $v$ and spacing $\Delta x$ in the steady state. In these models, traffic flow might be stable, metastable, or unstable.

---

[1] Note that Kerner and Konhäuser (1993, 1994) firstly proposed to use equilibrium speed density relationship with a turning point. Later, based on the empirical data available to him (Kerner and Rehborn, 1996, 1997; Kerner, 1998), Kerner believes that the theory is not able to correctly simulate real traffic and he proposed the three-phase theory (Kerner 2004, 2009).

Disturbances in the metastable and unstable traffic flows could grow and develop into jams via a subcritical Hopf bifurcation (Lee et al., 1998; Helbing et al., 1999).

In these theories, the traffic flow is classified into free flow and congested flow states. Therefore, these models are named as two-phase models by Kerner (2013). In contrast, based on empirical data available to him (Kerner and Rehborn, 1996, 1997; Kerner, 1998, 2004, 2009), Kerner claimed that the congested flow can be further classified into synchronized flow and wide moving jam. Usually phase transitions from free flow to synchronized flow occur firstly. At the upstream of the synchronized flow, there exists a pinch region, in which pinch effect induces small jams. The small jams propagate towards upstream, grow, merge, and develop into wide moving jams. In the three-phase traffic flow theory proposed by Kerner, it was supposed that the steady state of synchronized flow occupies a two-dimensional region in the flow-density plane, and the phase transitions are argued to be caused by the discontinuous characteristic of the probability of over-acceleration rather than the instability mechanism in traditional traffic flow models,

In recognizing the differences in the phase trajectories between accelerating vehicle streams and decelerating vehicle streams, Newell (1965) and later Zhang and Kim (2005) explicitly modeled transitions between acceleration, deceleration and cruising with different branches of speed-spacing or flow-density diagrams, which also spans a 2D region in the phase plane. But they do not distinguish between synchronized flow and wide moving jams. On the other hand, Kim and Zhang (2008) adopted a stochastic approach to model the wide scatter in congested flow, and derived the formula of the expected speeds of congestion waves. The latter shows remarkable consistency between computed and observed wave speeds for several observed waves.

Nevertheless, there is an ongoing debate between the two-phase and three-phase traffic flow theories, each side citing empirical evidence to support their arguments. The main difficulty to resolve this debate is a lack of high-fidelity traffic data that cover congestion from its birth to its death without gaps of information. Most of the data currently available are collected by point sensors (e.g. loop detectors), and distance between two neighboring sensors is usually 1-3 km for most freeway sections showing recurrent congestion patterns. As claimed by Treiber et al. (2010), "this is of the same order of magnitude as typical wavelengths of non-homogeneous congestion patterns" and therefore definitely leads to a loss of information pertaining the finer features of traffic flow. Daganzo et al. (1999) also pointed out that "no empirical studies to date describe the complete evolution of a disturbance" and thus we cannot "explain the genesis of the disturbances or the rate at which they grow".

The traffic community has also realized the importance of precise traffic data. As Schönhof and Helbing (2009) claimed, in order to decide the true cause of the spatiotemporal organization of traffic flow, which is the most central question of traffic theory, it is necessary to have spatiotemporal trajectory data of individual vehicles. Studies of this kind could be expensive and technologically challenging. For example, Knoop et al. (2008) have collected empirical data from a helicopter, using a digital camera gathering high resolution monochrome images. However, the images cover only 280 m of the motorway and due to the limited stability of the helicopter, only 210 m of the motorway could be used for vehicle detection and tracking. Similarly, the NGSIM-data cover only several hundred meters of the highway, which is also very limited. Moreover, the empirical observations via cameras are quite site-specific and contain many confounding factors (e.g., geometry such as on-ramps, off-ramps, and weaving area, bottleneck strength, traffic flow composition), which prevents us from forming a comprehensive understanding of traffic flow evolution. As a result, although there are vast amounts of empirical data, we are unfortunately still unable to understand how traffic dynamics evolve even in very simple scenarios, for example, a platoon of cars without overtaking is led by a car transitioning from one constant velocity to another.

Recently, we have reported an experimental study of car following behavior in a 25-car-platoon on an open road section (Jiang et al., 2014). The location and velocity of each individual car have been recorded by high precision GPS devices. Thus, the complete evolution of the disturbances (including their genesis and their growth rate) has been obtained. In the traffic experiment, we have changed the speed of the leading car so that various traffic flow situations have been studied, which goes beyond in-situ observations of traffic flow on roads.

It has been found that the spacing between a leading car and a following car can change significantly even though the speeds of the two cars are essentially constant and the velocity difference is very small. This feature clearly contradicts the fundamental assumption that there is a unique relationship between vehicle speed and its spacing in traditional car-following models. We have proposed two possible mechanisms to produce this feature. (i) In a certain range of spacing, drivers are not so sensitive to the changes in spacing when the velocity differences between cars are small. Only when the spacing is large (small) enough, will they accelerate (decelerate) to decrease (increase) the spacing. (ii) At a given velocity, drivers do not have a fixed preferred spacing. Instead they change their preferred spacing either intentionally or unintentionally from time to time in the driving process, see also the work of Wagner (2012, 2011, 2006) and Zhang and Kim (2008). Note that this observation can also be modelled by time-correlated perception errors of human drivers. Particularly, an error of the perceived gap leads to temporal changes of the realized time gap even for time-independent speeds and this is another way to reproduce the observations, see e.g., Chapter 12.3 of the book of Treiber and Kesting (2013).

We have proposed car following models based on the mechanism (ii). It has been shown that the growth pattern of disturbances has changed qualitatively and becomes qualitatively or even quantitatively consistent with that observed in the experiment. In contrast, the growth pattern of disturbances in the two-phase models such as the OV model, the FVD model, and the ID model is qualitatively different from the experimental findings.

This paper makes a further analysis of the experiments and reports new experimental findings. We also propose a car-following model based on mechanism (i) mentioned above and show that the model can reproduce the experimental results well.

The paper is organized as follows. In the next section, the experimental setup and previous experimental results are reviewed, and new experimental results are reported. Section 3 performs simulations of the GM models and the Gipps model, and shows that the simulation results are qualitatively different from the experimental ones. Section 4 proposes a new car-following model, the simulation results of which are well consistent with the experimental ones. Finally, Conclusions are given in section 5.

## 2. Experimental setup and results

*2.1. Experimental setup*

The experiment was carried out on January 19, 2013 on a 3.2 km stretch of the Chuangxin Avenue in a suburban area in Hefei City, China. See Figure 1(a) for the map of the section. There are no traffic lights on the road section. Since the road is located in a suburban area and has at least three lanes in each direction, there is no interference from other vehicles that are not part of the experiment.

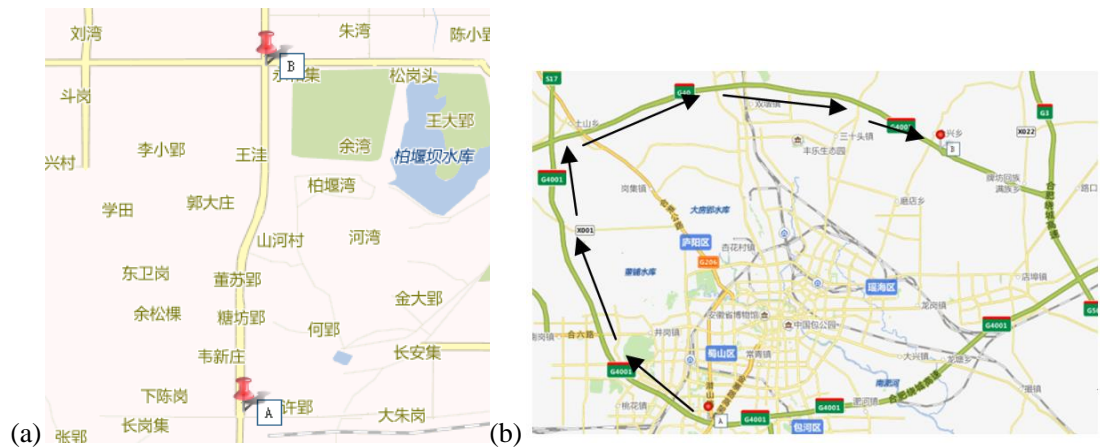

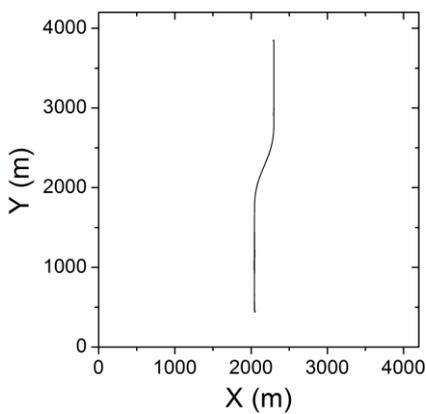  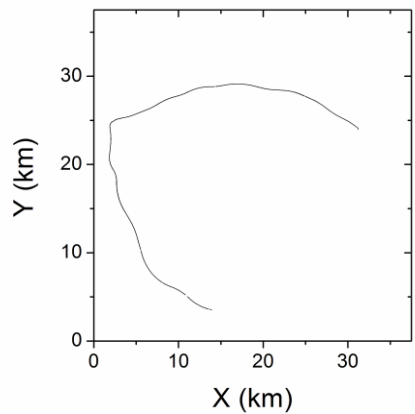

(c) (d)

Figure 1. (a) Map of the 3.2 km long experiment road section for the 25-car-platoon experiment, which is between points A and B. (b) Map of the experiment section for the 3-car-platoon experiment in the ring highway of Hefei City, which is from point A to B as shown by the arrows. (c) and (d) show trajectories of cars measured by the GPS.

High-precision difference GPS devices using MSAS (Multi-functional Satellite Augmentation System, a product from Japan) were installed on all of the cars to record their locations and velocities every 0.1 second. The measurement errors of the GPS devices were within ± 1 m for location and within ± 1 km/h for velocity. Once the experiment starts, the driver of the leading car is asked to drive the car at certain pre-determined constant velocity. Other drivers in the experiment are required to drive their cars as they normally do, following each other without overtaking. In two of the runs, the leading driver is asked to firstly drive his car at 15 km/h, then to drive the car without stepping on the gas pedal (the velocity is about 7 km/h in this case), and then to drive the car at 20 km/h. In other runs, the leading driver is asked to accelerate his car to a pre-determined velocity and then maintain the car at this constant velocity as accurate as he can.

The leading car is equipped with a cruise control system, which could be switched on when its velocity shown by the speedometer reaches 45 km/h. As a result, the fluctuation of the leading car is very small when its velocity is set at a value greater than or equal to 45 km/h. We note that for safety reason, the actual velocity of a car is lower than that shown by the speedometer, in particular when the velocity is high. See Figure 2 for the examples of the velocity of the leading car. When reaching the end of the road section, the car platoon decelerates, makes U-turn, and stops.[2] When all the cars have stopped, a new run of the experiment begins. We have carried out two sets of experiments, in which the sequence of the cars has been changed, see Table 1. Details of the runs in each set of the experiment are also presented in Table 1.

Videos for two typical scenarios of the experiment can be downloaded freely from http://www.plosone.org/article/info%3Adoi%2F10.1371%2Fjournal.pone.0094351.

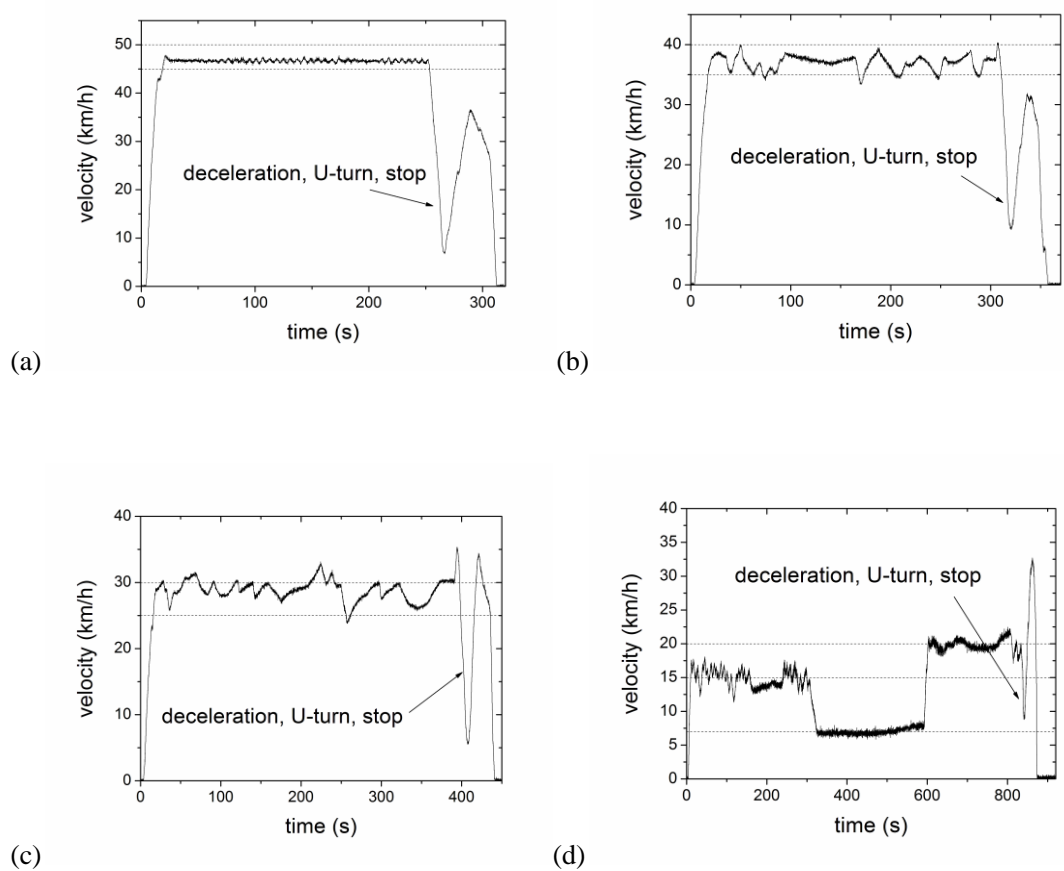

Figure 2. The velocity of the leading car. The driver of the leading car is asked to drive at (a) 50 km/h. (b) 40 km/h. (c) 30 km/h. (d) firstly at 15 km/h, then to drive the car without pressing the accelerator pedal (the velocity is about 7 km/h in this case), and then to drive the car at 20 km/h. Note that for safety reason, the actual velocity of a car is lower than that shown by the speedometer, in particular when the velocity is high. The dashed lines are added visual guides to help read the speeds on the graphs.

---

[2]Between the first run and the second run in the first set of experiment, the platoon has not stopped after the U-turn. The first run in the first set of experiment is regarded as a test run, and the data have not been used.

Table 1 Details of the experiments.

| Sequence of Car in the platoon | Car No. | | Run No. in the experiment | Velocity of leading car (km/h) | |
|---|---|---|---|---|---|
| | 1st set of experiment | 2nd set of experiment | | 1st set of experiment (12 runs) | 2nd set of experiment (23 runs) |
| 1 | 22 | 22 | 1 | 50 | 50 |
| 2 | 1 | 11 | 2 | 20 | 15-7-20 |
| 3 | 2 | 12 | 3 | 45 | 55 |
| 4 | 3 | 13 | 4 | 30 | 40 |
| 5 | 4 | 15 | 5 | 55 | 60 |
| 6 | 5 | 16 | 6 | 40 | 35 |
| 7 | 6 | 17 | 7 | 55 | 45 |
| 8 | 7 | 18 | 8 | 35 | 30 |
| 9 | 8 | 19 | 9 | 60 | 50 |
| 10 | 9 | 20 | 10 | 25 | 25 |
| 11 | 10 | 21 | 11 | 45 | 55 |
| 12 | 11 | 23 | 12 | 30 | 40 |
| 13 | 12 | 25 | 13 | | 50 |
| 14 | 13 | 1 | 14 | | 15-7-20 |
| 15 | 14 | 2 | 15 | | 55 |
| 16 | 15 | 3 | 16 | | 40 |
| 17 | 16 | 4 | 17 | | 60 |
| 18 | 17 | 5 | 18 | | 35 |
| 19 | 18 | 6 | 19 | | 45 |
| 20 | 19 | 7 | 20 | | 30 |
| 21 | 20 | 8 | 21 | | 50 |
| 22 | 21 | 9 | 22 | | 25 |
| 23 | 23 | 10 | 23 | | 55 |
| 24 | 24 | 14 | | | |
| 25 | 25 | 24 | | | |

*2.2 Previous experimental results*

In our previous paper (Jiang et al., 2014), we have shown that:

(i) Even if the preceding car moves with an essentially constant speed, the spacing of the following car still fluctuates significantly, see Fig.3.

(ii) Stripe structure has been observed in the spatiotemporal evolution of traffic flow, which corresponds to the formation and development of oscillations. When traffic flow speed is low, cars in the rear part of the 25-car-platoon will move in a stop-and-go pattern, see left panels in Fig.4.

(iii) The standard deviation $\sigma_v$ of the time series of the velocity of each car increases along the platoon in a concave or linear way, see Fig.5. However, the physical limits of speeds implies that if we had a much longer platoon, the variations of speed of cars in the tail of the platoon would be capped and the line would bend downward, making the overall curve concave shaped.

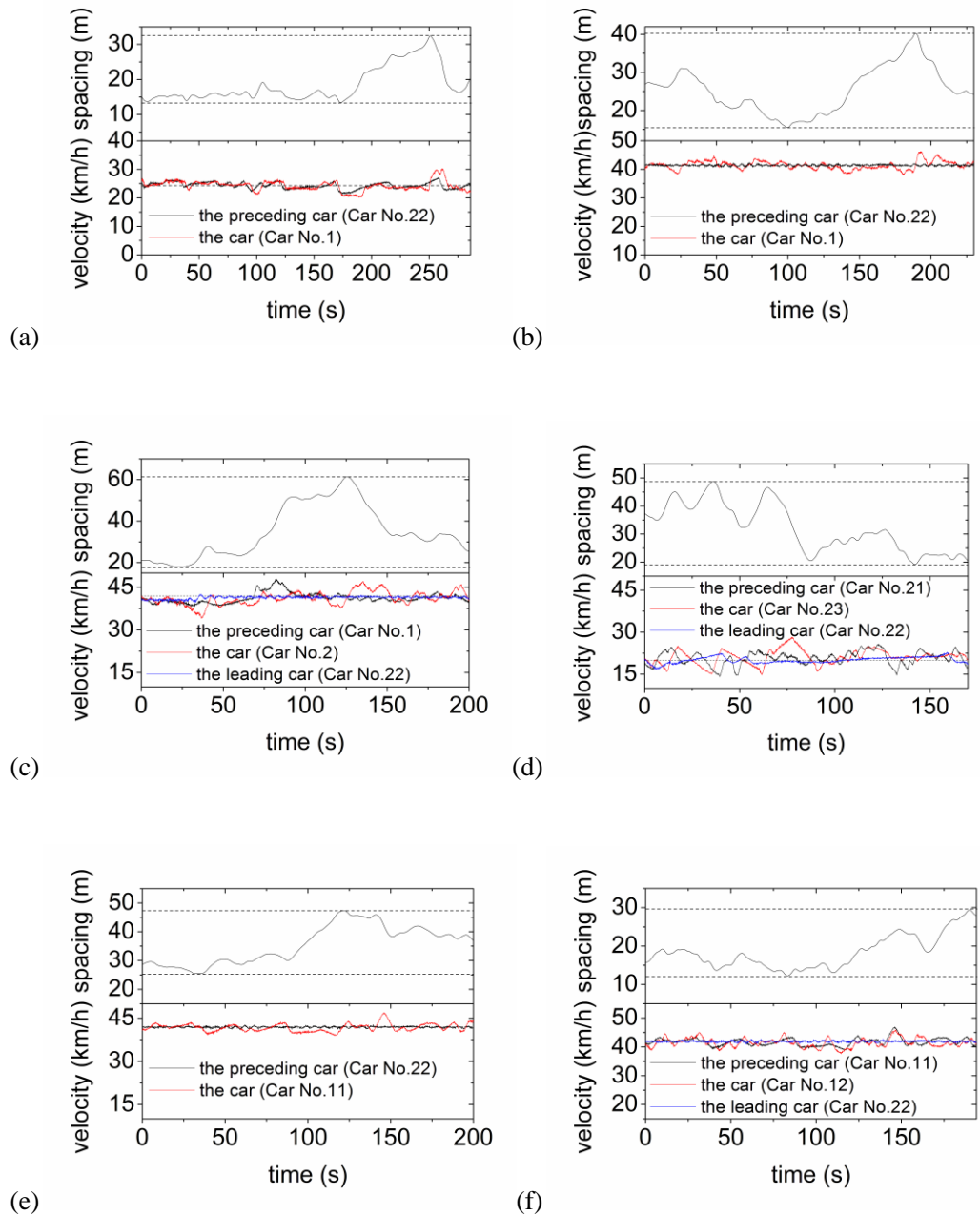

Figure 3. Evolution of the spacing of a car as well as the velocities of the car and its preceding car. (a) and (b), Car No.1 (the 2nd car in the platoon) in two different runs in the 1st set of experiment. (c) Car No.2 (the 3rd car in the platoon) in the 1st set of experiment. (d) Car No.23 (the 23rd car in the platoon) in the 1st set of experiment. (e) Car No.11 (the 2nd car in the platoon) in the 2nd set of experiment. (f) Car No.12 (the 3rd car in the platoon) in the 2nd set of experiment.

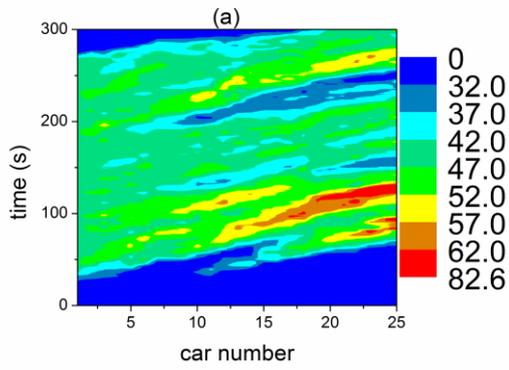
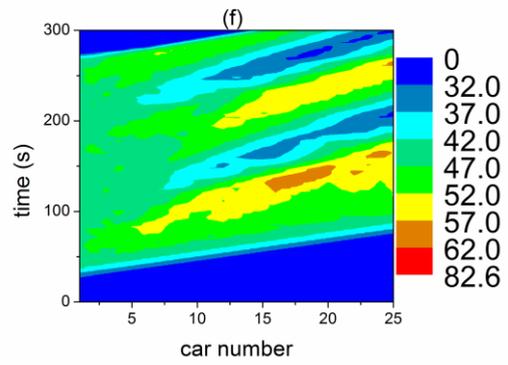
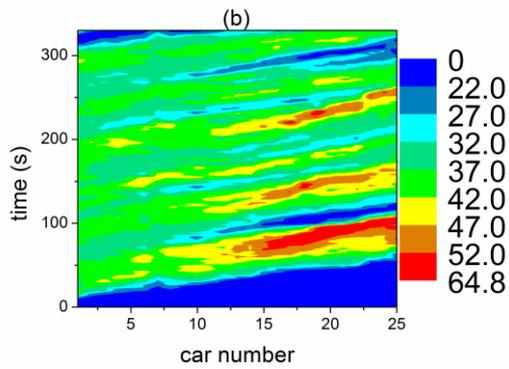
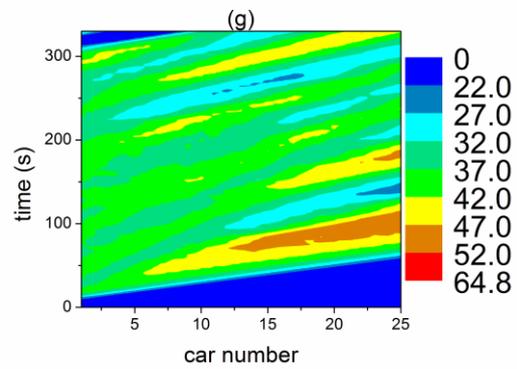
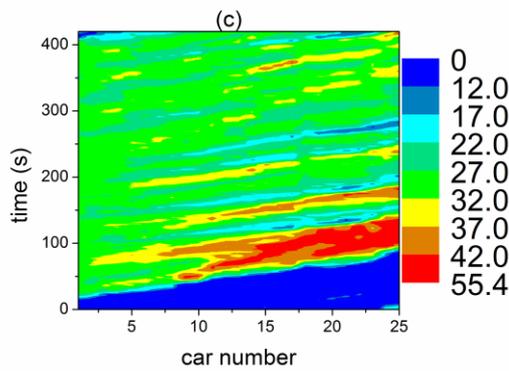
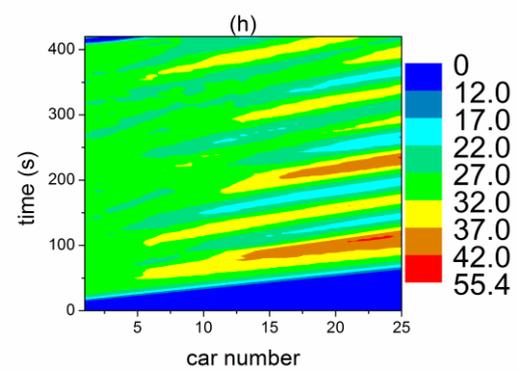

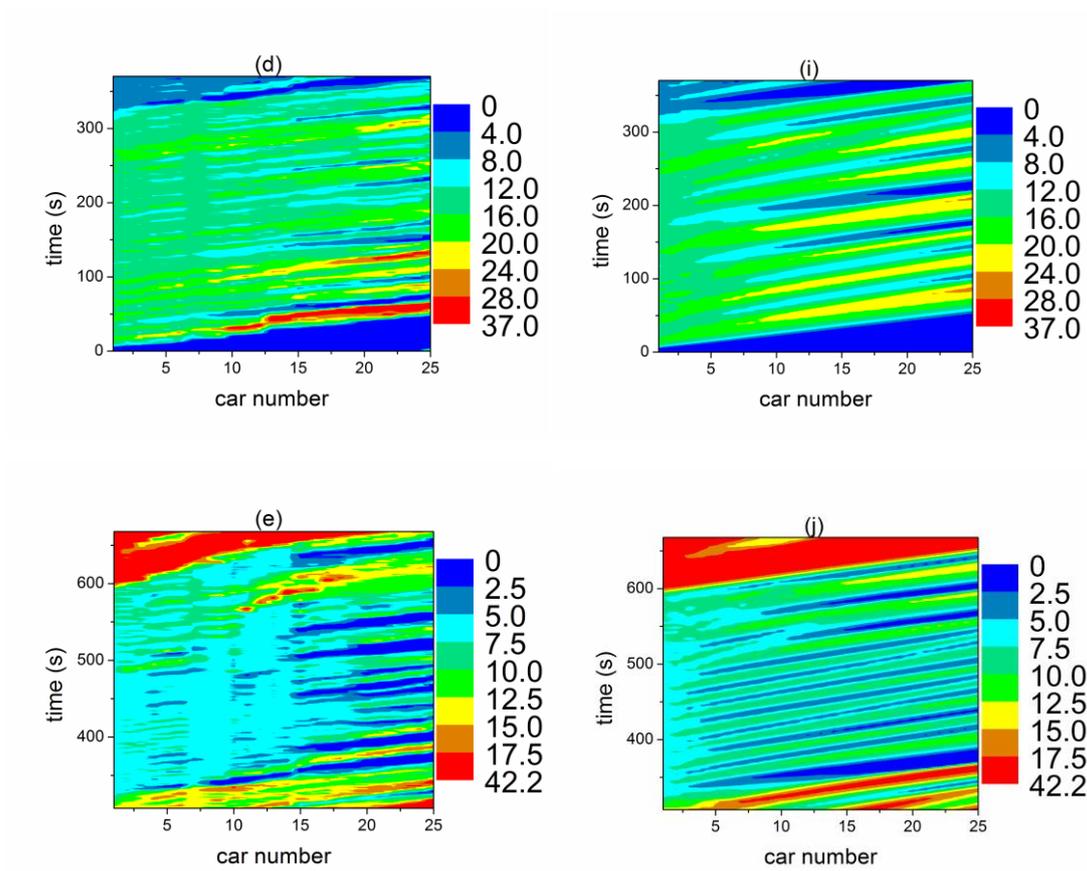

Figure 4 Evolution of the spatiotemporal pattern of car speed (unit km/h) in the experiment (left panels) and in the simulation by using the model proposed in section 4 (right panels). In the simulation, the speed of the leading car is set the same as in the experiment.

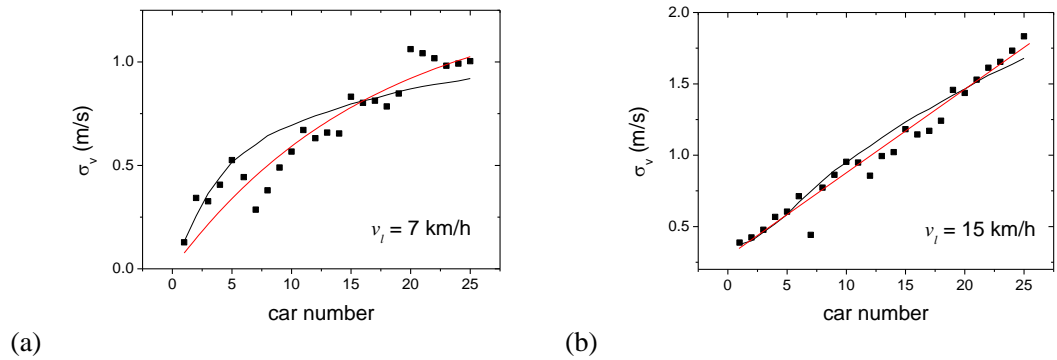

(a)                                                         (b)

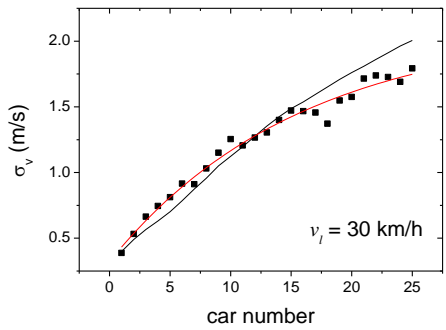
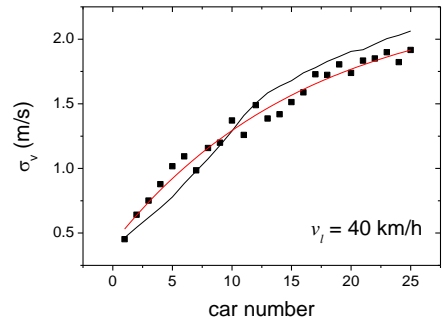

(c)                                                             (d)

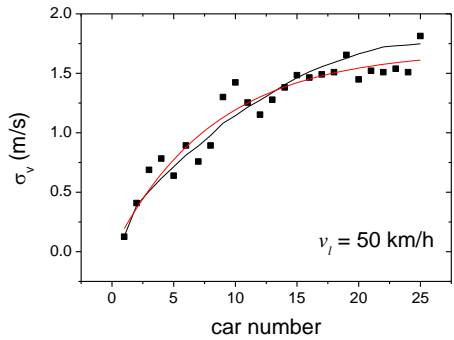

(e)

Figure 5    Comparison of the simulation results of the model proposed in section 4 (black solid lines) and experiment results (scattered data) of the standard deviation of the velocities of the cars. The red lines are fitted line.

*2.3 New experimental results*

     Figure 6 shows two time series of the spacing of car No.1 (the 2nd car in the platoon) as well as the velocities of this car and its preceding car in the first set of experiments. Figure 6(a) shows that the spacing changes significantly between 30 m and 60 m when the cars move with velocity around 51 km/h. However, Figure 6(b) shows that when the cars move with velocity around 56 km/h, the spacing fluctuates below 25 m, which is even smaller than the minimum spacing in Figure 6(a).

     Figure 7(a) shows the velocity and the spacing of car No.2 (the 15th car in the platoon) in two different runs in the second set of the 25-car experiment. The velocity of the car fluctuates between about 35 km/h and 58 km/h, and the average velocity equals to about 46 km/h in both runs. However, the average spacing of one run (red line) is much larger than the other run (black line). Figure 7(b) shows another similar example.

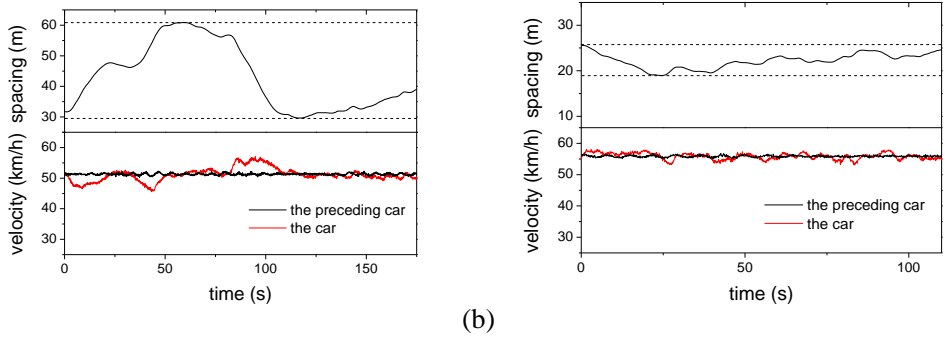

(a)                                                                                          (b)

Figure 6    The spacing of car No.1 (the 2nd car in the platoon) as well as the velocities of this car and its preceding car in the 1st set of experiment.

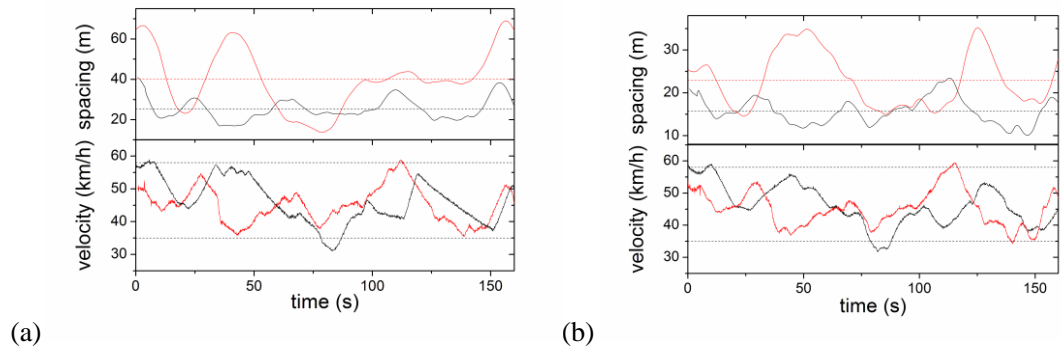

(a)                                                                                          (b)

Figure 7    The velocity and the spacing of car (a) No.2 (the 15th car in the platoon) (b) No.3 (the 16th car in the platoon) in two different runs in the second set of the 25-car experiment. The dashed lines in the top panels show the average spacing.

     We have also carried out another set of experiments, studying car-following behaviors in a 3-car-platoon that moves with high speed on the ring highway of Hefei City, see Figure 1(b). Since traffic is light in the experiment time period (from 9:55 am to 10:40 am) and the highway has three or four lanes in the experiment section, the platoon was not hindered by other vehicles and no vehicle has cut in the platoon during the experiment. The leading car in the experiment is asked to move with different constant velocity in different time intervals.

     Figure 8 shows the experimental results of the high speed experiment. The ratio $\sigma_v/v_{ave}$ and $\sigma_{\Delta x}/\Delta x_{ave}$ equal to 0.012 and 0.182 in panel (a), 0.0128 and 0.139 in panel (b), 0.048 and 0.242 in panel (c), 0.0426 and 0.205 in panel (d). Here $\sigma_v$ and $v_{ave}$ are standard deviation and average of velocity, $\sigma_{\Delta x}$ and $\Delta x_{ave}$ are standard deviation and average of spacing. These results show that the fluctuations in spacing are much larger than that in speeds.

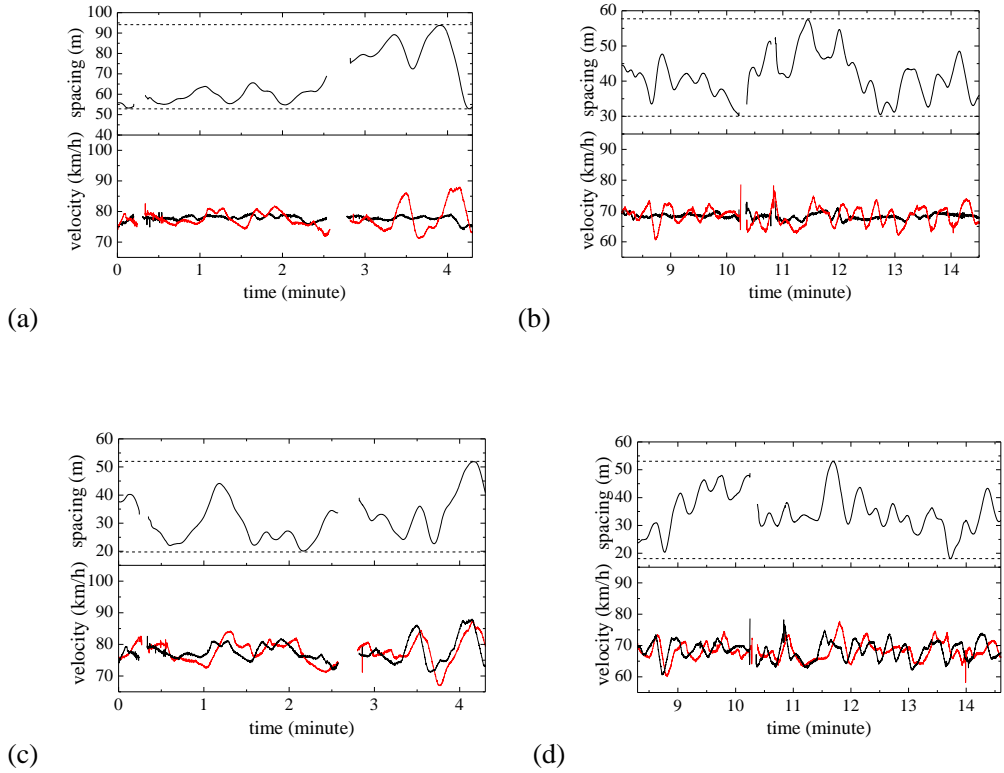

Figure 8 The velocity and the spacing of the car in the 3-car-platoon experiment on the Hefei ring highway. (a) and (b), the second car, (c) and (d), the third car. Red line corresponds to velocity of the car, and black line corresponds to velocity of preceding car. The gaps of the curves are due to temporary loss of GPS signal when cars move under bridges.

Next we present the velocity distributions of the cars in the 25-car-platoon experiment. In Fig. 9, the panels (a)-(e) show velocity distributions for different leading car velocities: (a) 50 km/h, (b) 40 km/h, (c) 30 km/h, (d) 15 km/h, (e) 7 km/h. The black squares are experimental results, the blue circles are simulation results of the model proposed in section 4 (to be described later in that section), and the red lines are the fitted Normal distribution curves. As shown in Fig.9, generally the velocity distributions can be well fitted by Normal distributions. There are two exceptions. One exception is when traffic is stop-and-go at low speeds, there appears another peak probability near zero speeds as shown by the red arrows in Figs.9(d) and 9(e). The other exception is when the leading car moves at low constant velocity (7 km/h) in panel (e), some drivers drive in a peculiar way: they did not step on the gas pedal just as the leading car, and their cars move with constant velocity for some periods of time (see Fig 10). As a result, there is another peak in the probability distribution at 7km/h, marked by the blue arrows in Fig.9(e). Consistent with the (concave) growth of $\sigma_v$ with respect to the position of a car in the platoon, the velocity distributions become flatter for cars further away from the platoon leader.

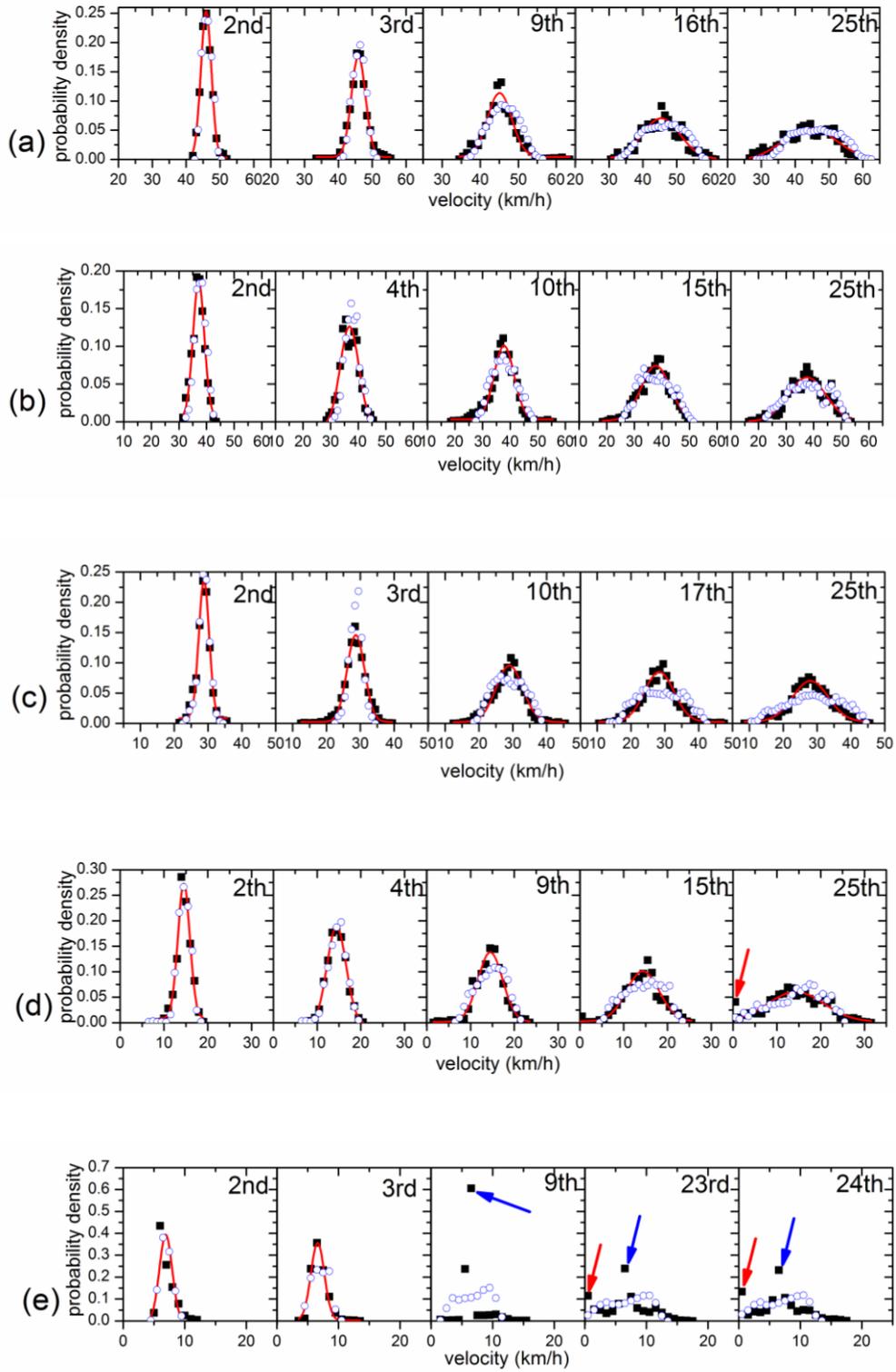

Figure 9. Velocity distribution of the cars and simulation fits.

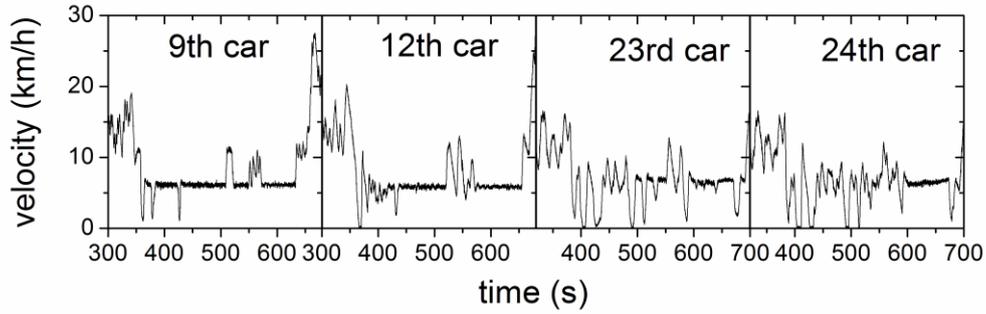

Figure 10  Examples of the velocity time series of some cars when traffic speed is low.

Now we study the platoon length L, which is related to the average spacing via s = L/(N-1), and the macroscopic density via ρ = 1/s, where N=25 is the number of cars in the platoon. Figs.11(a)-(d) show four examples, in which in different runs of the experiment, even if the leading car is asked to move with the same speed, the platoon length could be significantly different while the average velocity of all cars is approximately the same. Figure 9(e) shows an example, in which while the average velocity is smaller when the leading car moves with 20 km/h (compared with 25 km/h), the platoon length is larger. All the new experimental facts further demonstrate that traffic states span a 2D region in speed-spacing plane.

To show the relationship between individual spacing variation and the platoon length variation, we plot in Fig. 12 the spacing of cars and their differences in the second set of experiments where the leading car is asked to move with 60 km/h.  First, we plot in Fig.12(a) the spacing of cars in the first experimental run at two different time instants (indicated by blue and red arrows in Fig.11(a)): t=125 sec (black bars) and t=200 sec (red bars). We also plot the spacing of the cars at t =200 s in the second experimental run (indicated by green arrow in Fig.11(a)) in Fig. 12(a), which are shown by blue bars.  The difference in spacing between the two time instants in the first run are shown in Fig. 12(b), and that between the first and second run at t=200sec are shown in Fig. 12(c).  As can be seen from Fig. 12(b), the positive differences and the negative differences almost cancel out between the two time instants in the first run, so the platoon length remains relatively constant in this case (from 1121 m to 1054 m, a change of 6%). In the second case, the difference in spacing between the first run and the second run at time t=200sec are systematically positive, producing a large difference in platoon length (from 1121 m to 644 m, a 74% change). For the same average speed of travel, the spacing of cars in the second run is systematically lower than that in the first run, therefore its platoon length is significantly shorter, regardless of spacing distribution.

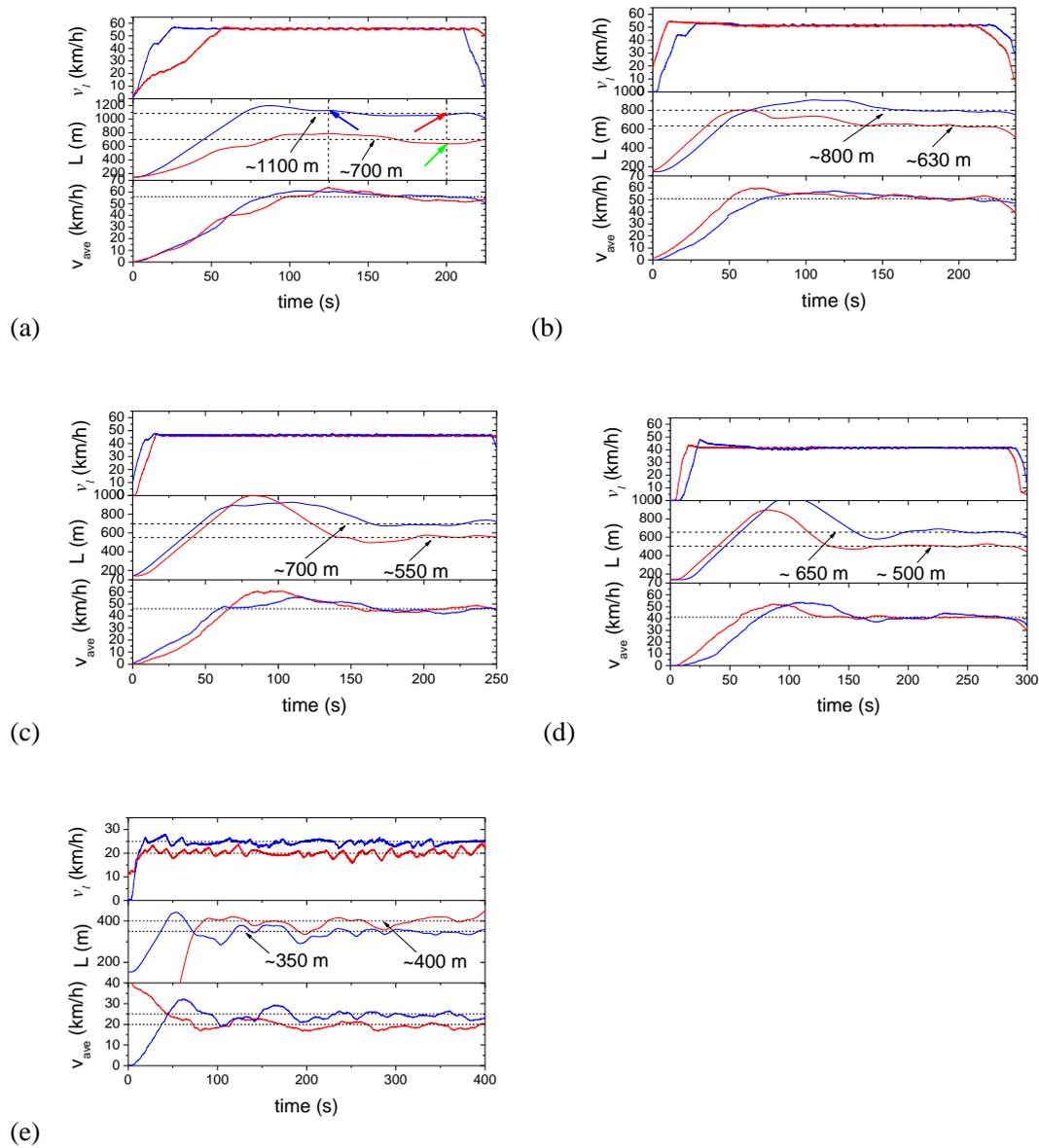

Figure 11 Evolution of the velocity of the leading car, the platoon length, and the average velocity of all cars in the platoon. The leading car is asked to move with speed (a) 60 km/h, (b) 55 km/h, (c) 50 km/h, (d) 45 km/h, (e) 25 km/h and 20 km/h. The blue and red color lines represent two runs of experiments with the same leading car speed in panels (a)-(d), and with different leading car speed in panel (e).

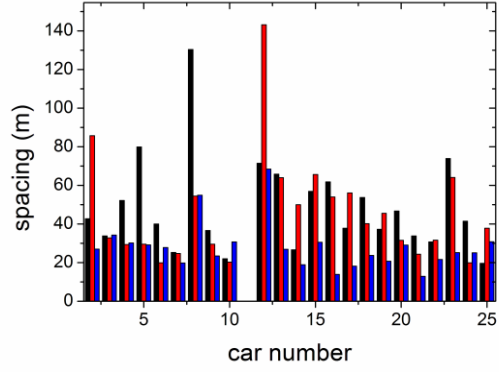

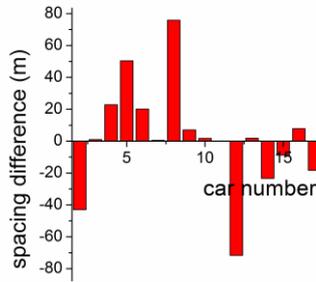 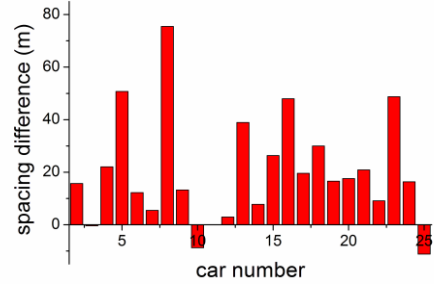

Figure 12 (a) The spacing of the cars in two different runs in the 2$^{nd}$ set of the experiment. Since the trajectory data of the 11$^{th}$ car are missing in one run of the experiment, the bars corresponding to the spacing of the 12$^{th}$ car show the spacing between the 10$^{th}$ car and the 12$^{th}$ car. Panels (b) and (c) shows the difference in spacing between the cars at t=150 sec and at t=200 sec in the first run, and at t=200 sec between the first and the second runs, respectively.

## 3. Simulation results of GM models and Gipps model

In previous work (Jiang et al., 2014), we have shown that in the two-phase models such as the OV model, the FVD model, and the ID model, the curve of $\sigma_v$ initially increases in a convex way, which is not consistent with the experimental results[3]. This implies that the instability mechanism is not likely correct in these models. In this section, we show that some other well known car-following models, namely the GM and the Gipps models, do not produce results consistent with the experiments.

In the following simulations, all cars are initially stopped with spacing $\Delta x$= 6 m. Then the leading car accelerates with a constant acceleration 1 m/s$^2$ until it reaches the velocity $v_l$. Then its speed begins to fluctuate with $v_l = v_l + \xi$, in which $\xi$ is a random number between $\pm 0.2$ m/s. In order to assure that there is no collision and the velocity does not exceed $v_{\max}$, we set $0 \leq v \leq v_{\max}$ and $v = 0$ if $\Delta x \leq 6$.

*3.1 simulation results of GM models*

---

[3]Li et al. (2014) have proved that in the Newell model, the disturbances grow in a convex way in the initial stage.

Figure 13(a) shows the simulation results of the GM model (1), in which the parameters used are $\lambda = 0.75 \text{s}^{-1}$, $\tau = 0.9$ s, $v_{\max} = 30$ m/s. One can see that the curve of $\sigma_v$ also increases in a convex way. When $v_l = 15$ km/h and 7 km/h, due to the lower bound of zero speed, the curve tends to become concave at the rear part of the platoon. One can image that if the platoon is longer, the curve will finally become concave due to the lower bound of zero speed and upper bound of maximum speed, no matter what $v_l$ is.

Figure 13(b) shows the simulation results of the GM model (2), in which the parameters used are $\lambda = 12$ m/s, $\tau = 0.9$ s, $m = 0$, $l = 1$, $v_{\max} = 30$ m/s. It can be seen that when $v_l = 30$, 40, and 50 km/h, the platoon is quite stable. When $v_l = 15$ km/h and 7 km/h, the curve of $\sigma_v$ initially increases in a convex way and then tends to be concave due to lower bound of zero speed.

Figure 13(c) shows the simulation results of the GM model (2), in which the parameters used are $\lambda = 4.5$, $\tau = 0.9$ s, $m = 1$, $l = 1$, $v_{\max} = 30$ m/s. Note that with parameter $m = 1$, a stopped car will never start again. Therefore, we modify GM model (2) into

$$\frac{dv_n(t+\tau)}{dt} = \max(\lambda(v_n(t+\tau))^m / [x_{n-1}(t) - x_n(t)]^l, \lambda_1)[v_{n-1}(t) - v_n(t)]$$

(7)

and the parameter $\lambda_1 = 0.5 \text{ s}^{-1}$. One can see that when $v_l = 7, 15, 50$ km/h, the platoon is quite stable. When $v_l = 30$ and 40 km/h, the platoon is very unstable and the velocity of the second car has already fluctuated very strongly and unrealistically, see Figure 13(d).

The simulation results demonstrate that the GM models were not able to reproduce the observed experimental features of traffic flow.

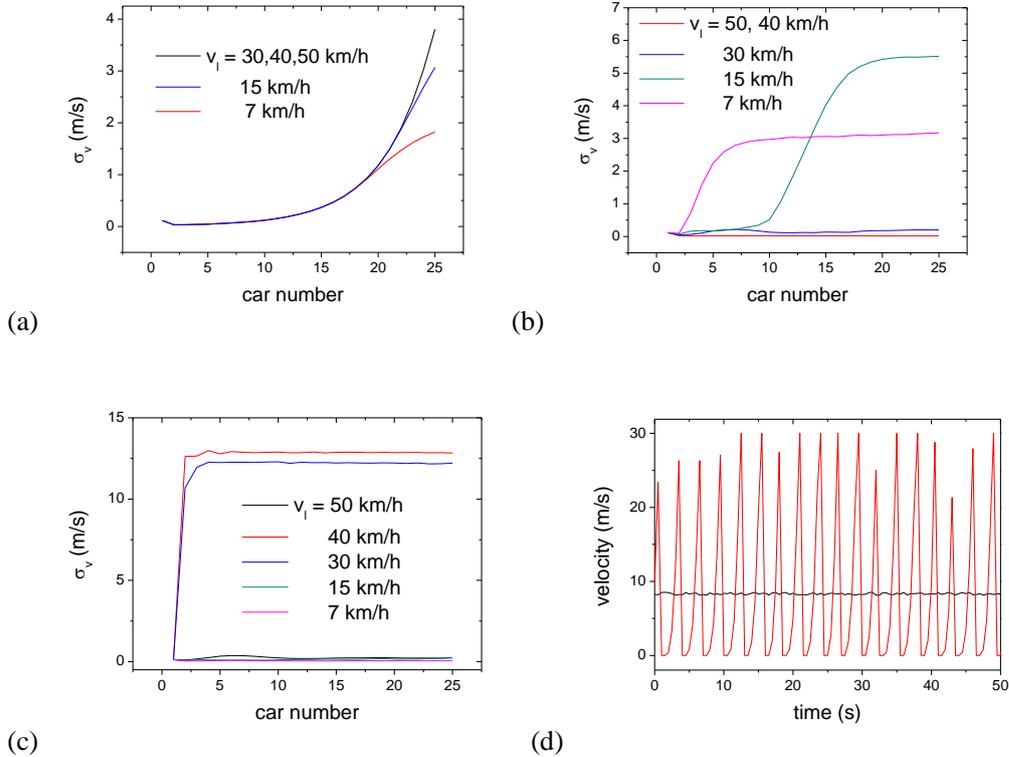

Figure 13  Simulation results of $\sigma_v$ in the GM models. (a) model (1), (b) and (c), model (2). (d) shows the velocity of the leading car (black line) and the second car (red line), corresponding to panel (c).

*3.2. Simulation results of the Gipps model*

In the Gipps model, the velocity of vehicle *n* is calculated by (Gipps, 1981)

$$v_n(t+\tau) = \min\left\{\begin{array}{l} v_n(t) + 2.5a\tau(1-v_n(t)/v_{max})\sqrt{0.025+v_n(t)/v_{max}}, \\ b\tau + \sqrt{b^2\tau^2 - b[2(x_{n-1}(t)-x_n(t)-S) - v_n(t)\tau - v_{n-1}^2(t)/\hat{b}]} \end{array}\right\} \quad (8)$$

In the simulations, the parameters are $\tau = 2/3$ s, $a = 2$ m/s², $b = -3$ m/s², $\hat{b} = -2.5$ m/s², $S = 6$ m, $v_{max} = 30$ m/s.

Figure 14 shows the simulation results. When $v_l = 7$ and 15 km/h, the platoon is quite stable. When $v_l = 30$, 40, and 50 km/h, the platoon is unstable and the curve of $\boldsymbol{\sigma_v}$ increases in a convex way. The feature of Gipps model is thus not consistent with the experimental results, either.

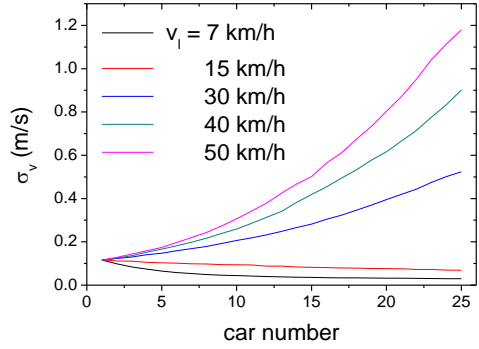

Figure 14 Simulation results of $\boldsymbol{\sigma_v}$ in the Gipps model

## 4. A new model and simulation results

In this section we propose a car-following model based on the assumption that in certain ranges of speed and spacing, drivers are insensitive to the changes in spacing when the velocity differences between cars are small.

It is supposed that there is a two-dimensional region R in the velocity-spacing plane in which car *i* moves as follows. If the absolute value of the velocity difference with the preceding car $i-1$ is smaller than a threshold, i.e., $|\Delta v_i| < \Delta v_c$, then the driver is not sensitive to either the velocity difference or the spacing, and he/she does not wish to change the speed. However, as it is not possible for the driver to control the accelerator pedal exactly to always maintain a constant velocity, the acceleration of the car changes randomly from $a(t)$ to[4]

$$a(t+\Delta t) = \max(\min(a(t)+\xi, 0.1), -0.1) \quad (9)$$

in each time step $\Delta t = 0.1$. Here $\xi$ is a uniform random acceleration within the range $[-0.02, 0.02]$. On the other hand, if $|\Delta v_i| > \Delta v_c$, then the driver becomes sensitive to the velocity

---

[4] All variables in this section are measured in SI units with time in seconds and space in meters, unless otherwise mentioned.

difference. In this case, the car moves as

$$\frac{dv_i}{dt} = \lambda \Delta v_i \tag{10}$$

Finally, if the state of the car is outside of the region R in the velocity-spacing plane, the driver becomes sensitive to both velocity difference and spacing. In this case, the car movement is modeled by the FVD model

$$\frac{dv_i}{dt} = \kappa[V(\Delta x_i) - v_i] + \lambda \Delta v_i \tag{11}$$

where V is the optimal velocity (OV) function, $\kappa$ and $\lambda$ are sensitivity parameters.

The two-dimensional region R is supposed to be bounded by five straight lines $v = 0.5(\Delta x - 6.8)$, $v = 0.22\Delta x + 5.5$, $v = \Delta x - 6$, $v = v_{max}$, and $v = 0$. The OV function used is $V(\Delta x) = \max(\min(v_{max}, 0.7(\Delta x - 6)), 0)$, see Figure 15. To avoid car collision, we set $v = 0$ if $\Delta x \leq 6$.

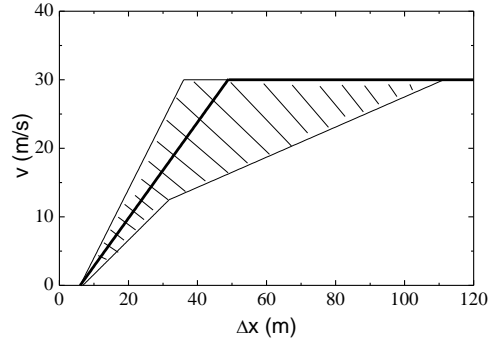

Figure 15   The two-dimensional region R (hatched) and the OV function (thick lines).

We would like to mention that Kerner firstly stipulated that traffic state should occupy a 2D region and proposed several microscopic models based on the assumption, such as the Kerner-Klenov model (Kerner and Klenov, 2002), and the Kerner-Klenov-Wolf cellular automaton model (Kerner et al., 2002). In these models, drivers always try to adapt to the velocity of the preceding cars, subject to stochastic acceleration and deceleration. Kerner's models, in particular the Kerner-Klenov model, were sometimes criticized due to their complex formulation of stochastic acceleration and deceleration. Comparing with Kerner's model, we have introduced a threshold $\Delta v_c$, beyond which the stochastic factors are neglected and below which the speed adaptation is neglected. Our formulation of stochastic factors is much simpler and easier to understand.

Now we present simulation results of the model. In the simulations, the parameters are set as $\kappa = 0.4$, $\lambda = 0.35$, $v_{max} = 30$, $\Delta v_c = \min(\max(0.6, 0.054v_i + 0.15), 1.0)$. As in the experiment, we consider a standing platoon of 25 cars. We let the leading car move as did the leading car in the experiment. The curves of the standard deviation of the velocities of the cars in the model are shown in Figure 5. One can see that the simulation results are not only qualitatively but also quantitatively in good agreement with the experimental ones. The right panels in Figures 4 show the spatiotemporal patterns of the velocities simulated from the model, in which formation of the stripes is also similar to those in the experiment. The speed distribution of the simulation results is compared with the experimental results in Fig.9. One can see that they are in good agreement except at velocity 7 km/h. This is partially because the peculiar driving behavior of some drivers as shown in Fig.10 has not been considered in the model.

Figure 16(a) shows an example of the evolution of the spacing of car No. 2 and the velocity of car No.2 (the following car) and No.1 (the preceding car). One can see that the velocity fluctuation is small, but the spacing fluctuation is large, which is consistent with the experimental results shown in Figures 3 and 8. Figure 16 (b) shows an example of the evolution of velocity and spacing of one car (No.20) in different runs. One can see that while the average velocity is almost the same, the spacing fluctuation as well as the average spacing is significantly different, which is in consistent with the experimental results shown in Figure 7.

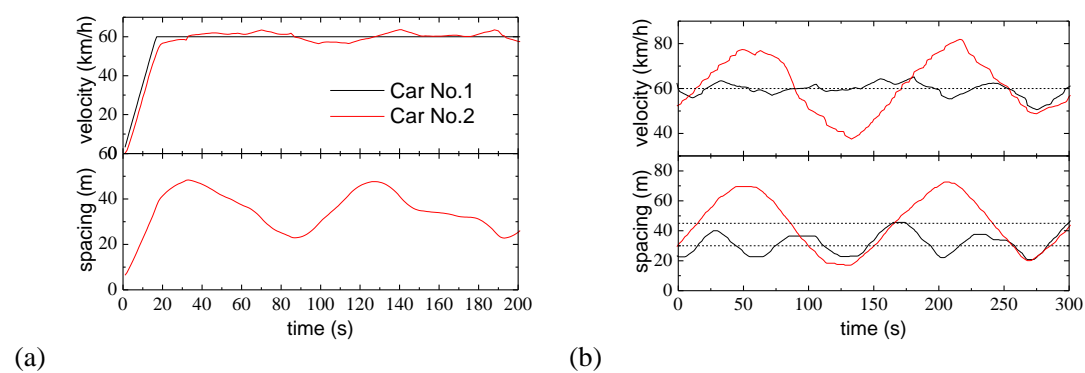

(a) (b)

Figure 16. (a) Simulation results of evolution of the spacing of car No. 2 and the velocity of car No.2 and No.1; (b) Simulation results of evolution of the velocity and spacing of car No. 20 in two different runs. The two dashed lines in the bottom panel show the average spacing of the two runs. The leading car moves with speed 60 km/h.

Figure 17 shows evolution of the platoon length, in which movement of the leading car is the same. As in the experiment, while the average velocity of the cars is essentially the same, the platoon length could be different under different runs.

However, the difference of platoon length is not so large as in experiment. Moreover, when the leading car velocity is small, the difference of platoon length under different runs is small. For example, simulation shows that the platoon length is always around 420 m when the leading car moves with speed 25 km/h, and the platoon length is always around 370 m when the leading car moves with speed 20 km/h. This is because, the 2D region becomes narrow with the decrease of speed. As a result, the phenomenon presented in Figure 11(e) has not been reproduced. One possible reason is that the heterogeneity of cars/drivers has not been considered in our simulations, which is clearly demonstrated in Figure 18. The driver of car No.3 tends to drive with spacing smaller than the velocity, while the driver of car No.10 tends to drive with spacing larger than the velocity.

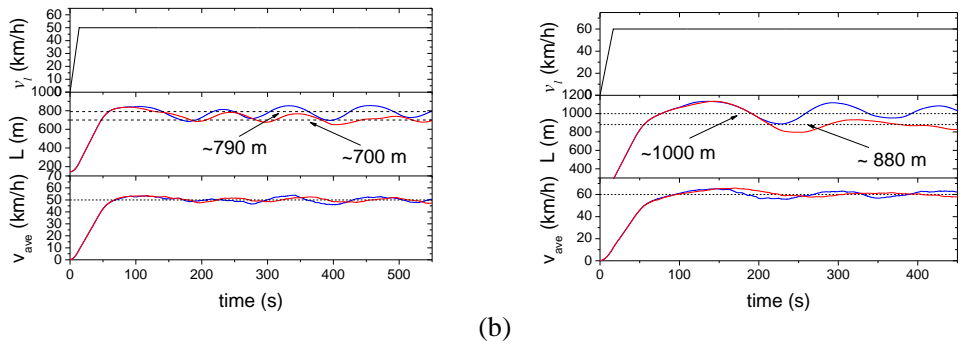

Figure 17  Simulation results of evolution of the velocity of the leading car, the platoon length, and the average velocity of all cars in the platoon. The leading car moves with velocity (a) 50 km/h, (b) 60 km/h.

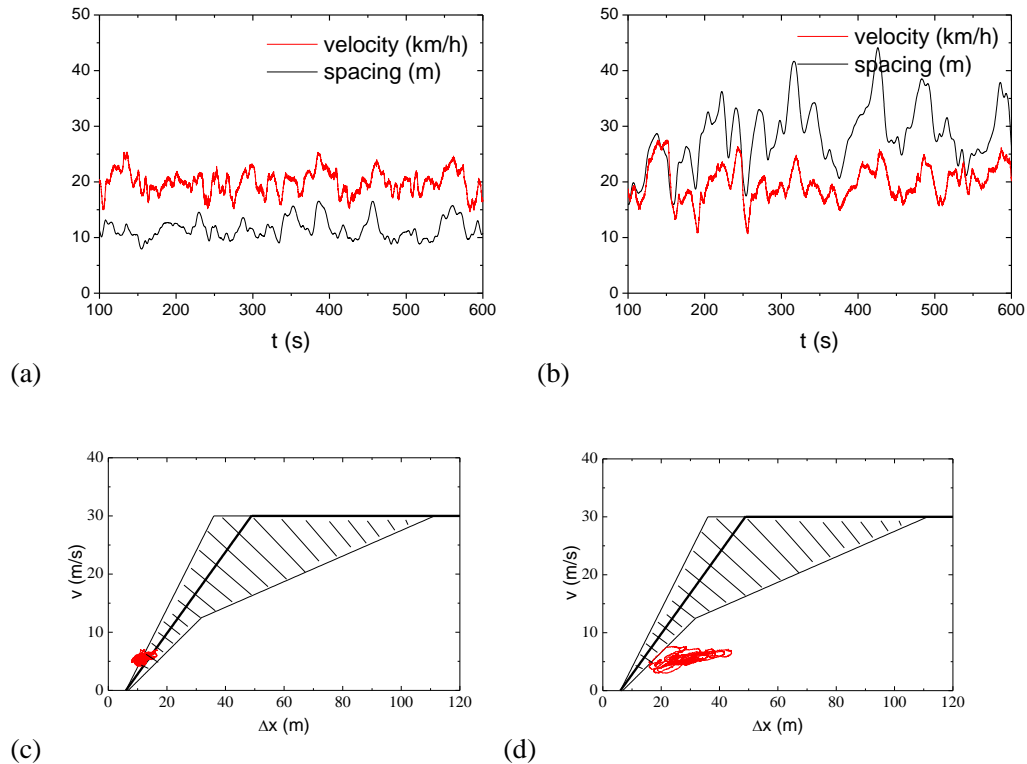

Figure 18  (a) and (b) show velocity and spacing of car No.3 and No.10, respectively, in which the leading car moves with velocity 20 km/h. (c) and (d) show the same data (red) in the spacing-velocity space. In (c) and (d), the background is the two-dimensional region R (hatched) and the OV function (thick lines) used in the simulations.

## 5. Conclusion

This paper makes a further analysis of our recent traffic experiments of car following behavior in a 25-car-platoon on an open road section. It has been shown that (i) the spacing could be smaller while the speed is larger (Figure 6); (ii) while the average speed is almost the same, the spacing fluctuation as well as the average spacing could be significantly different (Figure 7); (iii) the platoon length might be significantly different even if the average velocity of the platoon is essentially the same (Figure 11). We have also carried out another set of experiments, studying car-following behavior in a 3-car-platoon that moves with high speed on the highway, which shows similar results as in the 25-car-platoon experiment: while the velocities fluctuate slightly and the velocity difference is small, the spacing fluctuates in a wide range (Figure 8). The new experimental results thus further demonstrate that traffic states span a 2D region in the speed-spacing plane. Although the controversy about three-phase traffic theory is not settled with the present experimental data, it shows that any validate model should at least address this non-unique, one-to-many correspondence between speed and spacing, whether by adopting a three-phase diagram outright, or by considering driver heterogeneity or nonlinear, stochastic speed-spacing relations.

Next, we have shown that simulation results in the GM models and the Gipps model are inconsistent with the experimental results. Finally, we have proposed a car-following model based on the following mechanism. In a certain range of spacing, drivers are not so sensitive to the changes in spacing when the velocity differences between cars are small. Only when the spacing is large (small) enough, will they accelerate (decelerate) to decrease (increase) the spacing. The new model can reproduce the experimental results well.

In our future work, larger-scale experiments on longer road sections and with larger platoon sizes need to be carried out to study traffic flow evolution and to examine the proposed models. Actually, we have carried out a 50-car-platoon experiment (the leading speed is below 50 km/h) on Hefei old airport (which is closed) runway and taxiway (about 3 km each), and a 11-car-platoon experiment (the leading speed is above 60 km/h) on Hefei new airport highway (about 15 km). We will report the findings from these experiments in future publications.

**Acknowledgements**:
RJ was supported by the National Basic Research Program of China under Grant No.2012CB725404, the Natural Science Foundation of China under Grant No. 11422221 and 71371175, the Research Fund for the Doctoral Program of Ministry of Education of China under Grant No. 20123402110060. MBH was supported by Natural Science Foundation of China under Grant No.71171185. BJ was supported by the Natural Science Foundation of China under Grant No.71222101.